\documentclass[preprint,prc,aps,epsfig]{revtex4}
\usepackage{epsfig}
\def\be{\begin{equation}}
\def\ee{\end{equation}}
\def\bqa{\begin{eqnarray}}
\def\eqa{\end{eqnarray}}

\def\d3#1{\frac{d^3 p_#1}{(2\pi)^3 \; 2E_#1}}

\def\<{\left\langle}
\def\>{\right\rangle}

\begin{document}

\begin{center}
%%%%%%%% TITULO %%%%%%
{\LARGE \bf Elastic scattering and the proton form factor}

\vskip 1.0cm
                 
{\large F.\ Carvalho, 
A.\ A.\ Natale and C.\ M.\ Zanetti} \\[.2cm]
 
{\it Instituto de F\'{\i}sica Te\'orica \\
Universidade Estadual Paulista \\
Rua Pamplona, 145 \\
01405-900, S\~ao Paulo - SP \\
Brazil} 

\vskip 1.0cm

\end{center}

{\bf Abstract}

%--------------------------------------
We compute the differential and the total  cross sections for $pp$ scattering using the QCD pomeron model proposed by Landshoff and Nachtmann. This model is quite dependent on the experimental electromagnetic form factor, and it is not totally clear why this form factor gives good results even at moderate transferred momentum.  We exchange the eletromagnetic form factor by the asymptotic QCD proton form factor determined by Brodsky and Lepage (BL) plus a prescription for its low energy behavior dictated by the existence of a dynamically generated gluon mass. We fit the data with this QCD inspired form factor and a value for the dynamical gluon mass consistent with the ones determined in the literature. Our results also  provide a new determination of the proton wave function at the origin, which appears in the BL form factor. 
%--------------------------------------------------

\vskip 1.5cm
%\pacs
{PACS number(s): 12.40.Nn, 13.75.Cs\textit{}, 13.85.Lg}

\newpage  

The increase of hadronic total cross sections was theoretically predicted many years ago \cite{cheng}
and this prediction has been accurately verified by experiment \cite{hagiwara}. At present the main
theoretical approaches to explain this behavior are the Regge pole model and the QCD-inspired models.

In the Regge pole model the increase of the total cross section with the energy is attributed to 
the exchange of a colorless state having 
the quantum numbers of the vacuum: the pomeron \cite{pomeron}. According to the Regge pole theory the hadronic total cross section at high energy ($s\gg |t|$), where $s$ and $t$ are the Mandelstam variables of the process, behaves as $\sigma_{tot} \propto s^{\alpha(0)-1}$.
$\alpha(0)$ is the intercept of the Regge trajectory $\alpha(t)=\alpha_0+\alpha^{\prime} t$, where  $\alpha_0=1+\epsilon_0$, $\epsilon_0 \approx 0.08$ and $\alpha^{\prime} = 0.25$ GeV$^{-2}$. This behavior leads to a violation of the Froissart bound, although it is believed that the exchange of more pomerons eventually will lead to the unitarisation of the scattering amplitude.

Within Regge theory Donnachie and Landshoff \cite{dl} predicted the differential elastic proton-proton scattering to be given by

\begin{equation}
\frac{d\sigma}{dt}= const. \times F_1(t)^4 (\alpha^{\prime} s)^{2(\alpha (t) -1)}  \,\,\, ,
 \label{dsdt}
\end{equation}
where $F_1 (t)$ is the proton form factor measured in $ep$ elastic scattering. This expression fits the data
quite well at small $t$ \cite{cern}, and is reasonable even at moderate values of $t$. It is clear that this
cross section gives a non-trivial check for the functional form of $F_1(t)$, as well as it not clear at all
why the electromagnetic form factor works so well for the pomeron.

In this work we will consider a QCD pomeron model to compute Eq.(\ref{dsdt}). This model
has already been used successfully, therefore we use
the data in order to discuss its dependence on the
proton form factor. The idea is to exchange the eletromagnetic form factor by the asymptotic QCD proton form
factor determined by Brodsky and Lepage (BL) \cite{bl}. Of course this last one is not expected to be
suitable at the infrared (IR) region, but it is for this region that we introduce the concept of a dynamical
gluon mass and one hypothesis about how the form factor should behave in the IR, 
matching it to the BL proton form factor at high energy. Our model for the proton form factor is used to obtain the differential and the total cross-sections for $pp$ scattering, and it fits the data with a value for the dynamical gluon
mass consistent with the ones determined in the literature. In this calculation we
obtain a reasonable value for the proton wave function at the origin that appears in the BL form factor, which is used in many phenomenological applications, and turns out to be related to the dynamical gluon mass.

In the QCD framework the pomeron can be understood
as the exchange of at least two gluons in a color singlet state \cite{low}. A model
for the Pomeron has been put forward by Landshoff and Nachtmann where it is evidenced the importance of the QCD non-perturbative
vacuum \cite{landshoff}. One of the aspects of this non-perturbative physics appears as an infrared gluon
mass scale which regulates the divergent behavior of the Pomeron exchange \cite{halzen}.

The Landshoff and Nachtmann (LN) model can explain diffractive scattering data quite successfully \cite{halzen,outros}, at the same time that it is a quite simple model. The calculation of hadronic cross sections in this model are
straightforward, although there are two approximations that are usually performed in order to compare the
model to the data. First, the pomeron exchange dependence on the energy has to be introduced in an {\sl ad hoc} way. This means
that we multiply the scattering amplitude by a factor of the form $Z=(s/w_0^2)^{0.08+\alpha^{\prime} t}$, $w_0^2 \approx 1/\alpha^{\prime}$ is an energy scale.
We know that such behavior must come from the exchange of multiple gluon ladder diagrams described by the Balitski-Fadin-Kuraev-Lipatov (BFKL) equation \cite{bfkl}. However we also know that the exponent of the BFKL is not simply related to the data. Of course we expect that in the future the LN approach can be made compatible with the BFKL one. In the meantime we just follow as described above and introduce the factor $Z$ by hand. Secondly, the cross sections depend on the square of hadronic wave functions, which are ultimately related to their respective form factors.
As discussed above, in general it is the electromagnetic phenomenological form factor that is used in the actual diffractive cross section calculation \cite{halzen}. The ideal case would be the use of a proton form factor 
totally derived from QCD, and this is the point that will deserve our attention.

In the LN model the elastic pp differential cross section can be obtained from

\begin{equation}\label{dsigma}
  \frac{d\sigma}{dt}= \frac{|A(s,t)|^2}{16\pi s^2}
\end{equation}

\noindent
where the amplitude for elastic proton-proton scattering via
two-gluon exchange can be written as

\begin{equation}
A(s,t)= is8\alpha_{s}^2\left[T_{1} - T_{2}\right]
 \label{ampli}
\end{equation}

\noindent
with

\begin{equation}
T_{1}= \int\,d^{2}k
D\left(\frac{q}{2}+k \right)  D \left(\frac{q}{2}-k\right)|G_{p}(q,0)|^{2} \label{t1}
\end{equation}

\begin{eqnarray}
T_{2}&=& \int\,d^{2}k D\left(\frac{q}{2}+k\right)D\left(\frac{q}{2}-k\right)G_{p}
\left(q,k-\frac{q}{2}\right)  
\nonumber \\
&& \times \left[2G_{p}(q,0)-G_{p}\left(q,k-\frac{q}{2}\right)\right ]
\label{t2}
\end{eqnarray}

\noindent
where $G_p (q,k)$ is a convolution of proton wave functions

\begin{equation}
G_p(q,k)= \int \, d^2pd\kappa \psi^\ast(\kappa,p)\psi(\kappa,p-k-\kappa q).
\label{gdp}
\end{equation}
To estimate $G_p (q,k-q/2)$ it is usually assumed a proton wave function peaked at
$\kappa = 1/3$ implying that

\begin{equation}
 G_p\left(q,k-\frac{q}{2}\right)=
 F_{1}\left( q^2 + 9\left|k^2-\frac{q^2}{4}\right| \right).
\label{conv}
\end{equation}

\noindent
In the many calculations of this model up to now $G_p (q,0)$ was given by the Dirac form factor of the proton

\begin{equation}
F_{1}(t) = G_p(q,0)= \frac{4m^2 - 2.79t}{4m^2
-t}\frac{1}{(1-t/0.71)^2} \label{dirac} .
\end{equation}
This means that the strongly interacting pomeron sees the proton in the same way as it is seen by a photon.
This is not totally surprising, as discussed in the work of Landshoff and Nachtmann \cite{landshoff}, since the pomeron
formed by gluons that propagate in the vacuum up to a certain critical distance (the gluon mass in our case)
lead naturally at high energies to a pomeron that interacts as an isoscalar photon.  

In this work we would like to differ from the previous calculations exactly by the use
of the asymptotic form factor expression determined
by Lepage and Brodsky \cite{brodsky}, which is given by

\begin{equation}
F_{QCD}(t) = G_p(q,0) \approx \frac{C [\alpha_s(q^2)]^{2+4/3\beta}}{q^4}  \,\, ,
\label{qcd} 
\end{equation}
where $\beta =11-(2/3)n_{flavor}$ is the coefficient of the QCD $\beta$ function. The $q^4$ behavior comes from the
propagators of two gluons exchanged between the quarks forming the proton and $C$ is determined by the qqq wave function at the origin. Contrary to the electromagnetic form factor given by Eq.(\ref{dirac}) the one of Eq.(\ref{qcd}) is just the asymptotic one, 
and not normalized to $1$ at $q^2=0$.

We know that the BL proton form factor should be reliable at asymptotic energies, i.e.
it is just the tail of the actual form factor.
How can we perform a full QCD calculation of the differential $pp$ elastic scattering
without knowing the IR behavior of the form factor? To guess something about its 
IR functional form we must introduce in the form factor calculation the concept of a dynamically generated gluon mass,
which is already going to modify the expression of Eq.(\ref{qcd}).

Eq.(\ref{dsigma}) up to Eq.(\ref{conv}) were computed in Ref.\cite{halzen,outros}
using nonperturbative gluon
propagators endowed with a dynamical gluon mass, whose existence also imply
in an IR finite coupling constant \cite{natale}. The possibility
that QCD generates a dynamical gluon mass has been put forward by Cornwall \cite{cornwall},
and there are evidences for such behavior obtained from solutions of Schwinger-Dyson equations (SDE) \cite{cornwall,sde} and from lattice simulations (see Ref.\cite{port} and
references therein). These propagator and coupling constant have been used in
many phenomenological calculations that are sensible to their infrared finite behavior \cite{outros,mihara}, and are given by: 

\begin{equation} D_{\mu\nu}(q^2)= 
\left({\delta}_{\mu\nu}-\frac{q_{\mu}q_{\nu}}{q^2}\right)D(q^2),
\label{landau}
\end{equation}

\noindent
where the expression for $D(q^2)$, that was obtained by Cornwall as a fit to the numerical solution of a gauge invariant gluonic SDE, is equal to

\begin{equation}\label{propcorn}
 D^{-1}(q^2) = 
 \left[q^2 + M_g ^2(q^2)\right]bg^2\ln\left[\frac{q^2+ 4M^2_{g}}{\Lambda ^2} \right].
\end{equation}
where $M_g(q^2)$ is a dynamical gluon mass described by,
\be M^2_g(q^2) =m_g^2 \left[\frac{ \ln
\left(\frac{q^2+4{m_g}^2}{\Lambda ^2}\right) } {
\ln\left(\frac{4{m_g}^2}{\Lambda ^2}\right) }\right]^{- 12/11}
\label{mdyna} \ee
$\Lambda$($\equiv\Lambda_{QCD}$) is the QCD scale parameter. The infrared finite coupling constant obtained in
the same procedure has the following expression \cite{cornwall} 

\be \alpha_{s} (q^2)= \frac{4\pi}{\beta \ln\left[
(q^2 + 4M_g^2(q^2) )/\Lambda^2 \right]}, \label{acor} \ee
A typical phenomenological value for the dynamical gluon mass
is \cite{cornwall,halzen}
\begin {equation}
m_g = 500 \pm 200 \quad \mbox{MeV}
\label{mg}
\end{equation}
for $\Lambda = 300$ MeV. Finally, note that these solutions
have been obtained in the Euclidean space, as well as are the momenta in the above
equations.

In practice the SDE equations have to be solved with approximations and there
are different functional forms proposed for the gluon propagator. In particular,
a quite simple expression has been proposed in Ref.\cite{aguilar}

\begin{equation}\label{propac}
 D^{-1}(q^2) = \left[q^2 + M_g ^2(q^2)\right] \,\,\, ,
\end{equation}
and this expression, even neglecting the $q^2$ dependence in $M_g ^2(q^2)$, 
is the one that we shall use in our calculation. The same happens to the
IR behavior of the running coupling constant. Its behavior at $q^2=0$
is around $1$ \cite{mihara,deur}, and since the calculation is not strongly
dependent on the ultraviolet logarithmic behavior of the coupling constant,
we will just assume a constant value for this quantity. The effect of 
these approximations will be commented ahead.

Once we admit the idea of a dynamically generated gluon mass we can guess
what could be expected for the proton form factor. Asymptotically, if
we neglect the $\alpha_s$ dependence in Eq.(\ref{qcd}), we have

\begin{equation}
G_p(q,0)|_{q^2\rightarrow \infty} \approx \frac{C}{(q^2+m_g^2)^2}  \,\, .
\label{qcd2} 
\end{equation}
It could be argued that this approximation is rough due to the fact that we
neglected the dependence on $\alpha_s$. Actually this is not so trivial,
because Cornwall \cite{cornwall} points out that the product $g_s^2 D(q^2)$ is
independent of the coupling constant. In Ref.\cite{mihara} we have shown that for gluon masses in the
range of Eq.(\ref{mg}) the values of the infrared coupling constant vary between $0.5$ and $1$. We
will arbitrarily use the infrared value of $0.8$. Unfortunately we still have a poor knowledge
of the infrared QCD behavior, but if the Cornwall's argument is correct it may happens that our
approximation is much better than it looks like.

Since the gluon has its propagation limited by the value of the gluon mass ($m_g$), we expect that 
the IR proton form factor is approximately constant in a radius determined
by this mass, and falls off vary fast after that. The simplest choice that we can
make for this IR behavior is
\begin{equation}
G_p(q,0)|_{q^2\rightarrow 0} \approx \exp [{-q^2/2m_g^2}]  \,\, ,
\label{irff} 
\end{equation}
which is naturally normalized to $1$ at the origin. Note that more complicated expressions
could be used to describe the IR behavior, but they barely would reflect in such a simple way
a proton that is formed by the exchange of dynamically massive gluons.

The full form factor must be a combination of Eq.(\ref{qcd2}) and Eq.(\ref{irff}), and
must match at one intermediate scale, which we propose as being the dynamical gluon
mass scale, with the Brodsky and Lepage proton form factor. Therefore our ansatz for the proton form factor for the full range of
momenta is given by
\begin{equation}
G_p(q,0) =  \exp [{-q^2/2m_g^2}] \theta (2m_g^2 - q^2) +
\frac{9m_g^4/e}{(q^2+m_g^2)^2} \theta (q^2 -2m_g^2) \,\, .
\label{fullff} 
\end{equation}
Eq.(\ref{fullff}) is correctly normalized at the origin, match softly the low and
high energy of the proton form factor, and, if compared to Eq.(\ref{qcd2}), gives a
prediction for the proton wave function at the origin
\begin{equation}
C \equiv \frac{9m_g^4}{e} \,\,\, ,
\label{wave0} 
\end{equation} 
whose value will be determined when we compare our cross section calculation to the
experimental data.

We compute the differential pp scattering cross section in the LN model, as
performed in Ref.\cite{halzen}, with the help of the proton form factor of
Eq.(\ref{fullff}). In Fig.(1) the result is compared to the experimental data
from  {\sl et. al.} \cite{cern}. The data are well fitted assuming a dynamical gluon mass of $m_g\,=\,460$ MeV, for $\Lambda\,=\,300$ MeV. The calculations have some dependence on the gluon mass. We also present in Fig. (1) the results obtained using $m_g\,=\,400$ and $550$ MeV.
%-----------------------------------------------------
\begin{figure}[h]
\begin{center}
\centerline{\epsfig{figure=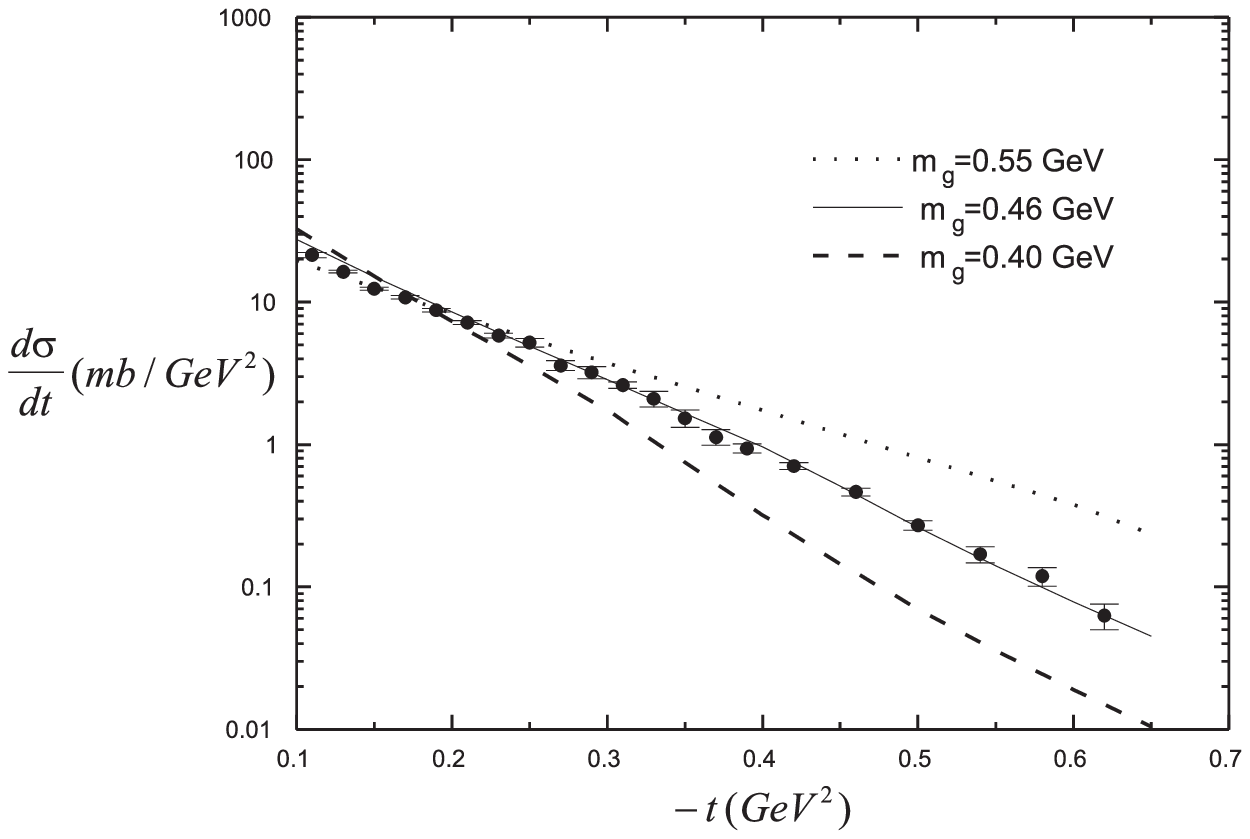,width=12cm}} %\,\,\,\,\,
\centerline{\small{Fig.(1) Differential pp scattering cross section from the LN model.}}
\end{center}
\end{figure}
We also compute the total cross section obtained from Eq.(\ref{ampli}) (after multiplication by the factor $Z$). It is clear that the LN model calculation, which turns out to be a function of $m_g$ and $C$, can reproduce only the Pomeron term (the one that increases with $s$) of the fit for the total cross sections proposed by Donnachie and Landshoff \cite{dl2}, where it is shown that the high energy behavior of $pp \,\, (p\bar{p})$ scattering is proportional to $21.7 s^{0.0808}$.

The high energy data on $pp$ scattering from Ref. \cite{pdg} is also shown in Fig.(2) and compared 
with our calculations, using the same values of the dynamical gluon mass, which turns out to be, together with the Donnachie and Landshoff fit, just another check for our results.

%-----------------------------------------------

%------------------------------------------
\begin{figure}[h]
\begin{center}
\centerline{\epsfig{figure=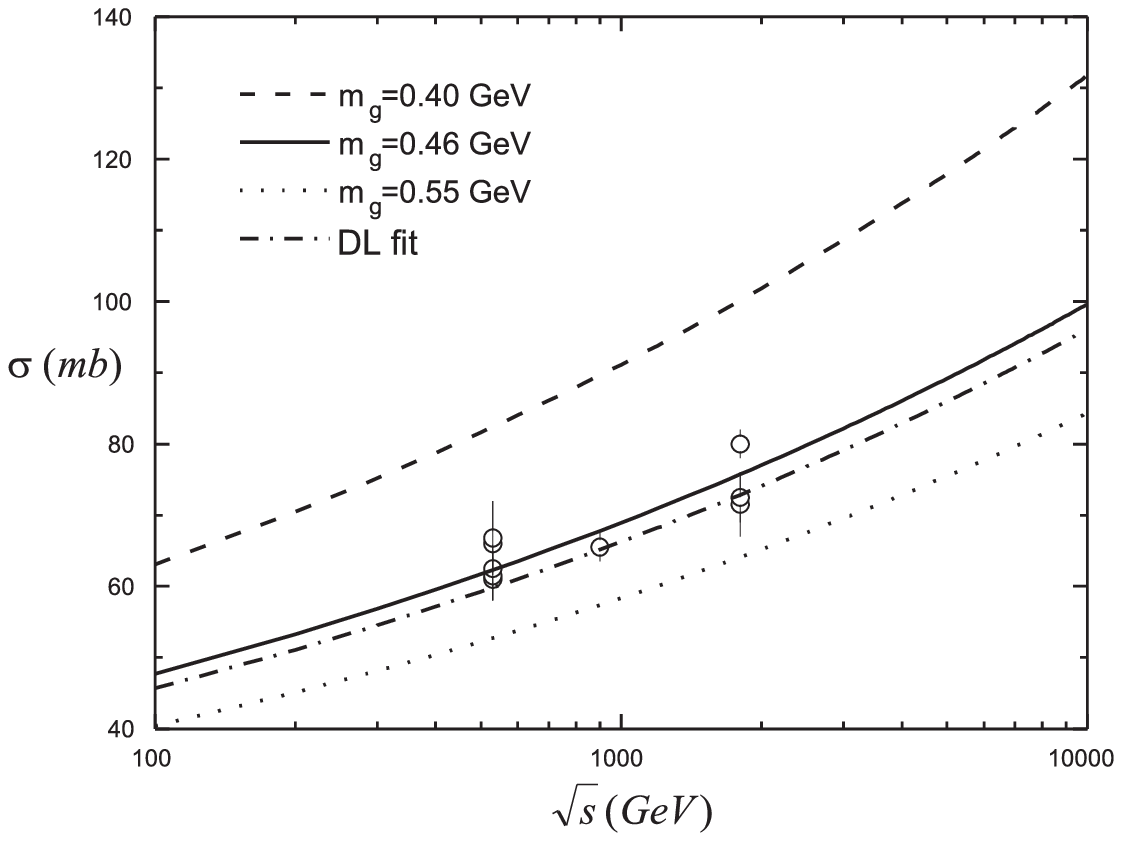,width=12cm}} %\,\,\,\,\,
\centerline{\small{Fig.(2) Proton-proton total cross section. The experimental data ($p \bar{p}$) are from Ref.\cite{pdg}.}}
\centerline{\small{ It is also shown the curve given by the Donnachie and Landshoff fit.}}
\end{center}
\end{figure}
From Fig. (1) and (2) we see that the data shows a preferred value $m_g\,=\,460$ MeV,
that is compatible with previous determinations of the dynamical gluon mass. With
this value we obtain
\begin{equation}
C= 0.15  \,\,\,\, {\textnormal GeV}^4 \,\,\, .
\label{cwave}
\end{equation} 

It is interesting that in Ref.\cite{halzen} the data of elastic proton-proton scattering is well fitted assuming a dynamical gluon mass of $m_g = 370 \quad {\textnormal MeV}$ for $\Lambda= 300 \quad {\textnormal MeV}$, while our
recent calculation in Ref.\cite{luna} shows excellent agreement with the data for $m_g = 400\,(-100)\,(+350) \quad {\textnormal MeV}$.
Our result is in agreement with such calculations and provides a new determination of the proton wave function at the origin. 

The value of $C$ is quite compatible with the results obtained from the branching ratio 
$\frac{\Gamma (\psi \rightarrow pp)}{\Gamma (\psi \rightarrow hadrons)}$, whose experimental value is $0.0022$.
In this calculation we use the expression for the branching ratio given by \cite{brodsky81}: 
\begin{equation}
\frac{\Gamma (\psi \rightarrow p\bar{p})}{\Gamma (\psi \rightarrow hadrons)}=  3.2\,10^6\,\alpha^3_s(s) \frac{|\vec{P}_{c.m.}|}{\sqrt{s}} \frac{<T>^2}{s^4}
\end{equation}
where $|\vec{P}_{c.m.}|/\sqrt{s} \simeq 0.4$, $s=9.6$ GeV$^2$ and
\begin{equation}
<T> \equiv \int_0^1 [dx] [dy] \frac{\phi^*(y_i,s)}{y_1 y_2 y_3}
\frac{x_1 y_3+x_3y_1}{[x_1 (1-y_1)+y_1(1-x_1)][x_3 (1-y_3)+y_3(1-x_3)]}
\frac{\phi^*(x_i,s)}{x_1 x_2 x_3}
\end{equation}
We performed the above calculation using 
\begin{equation}
\phi(x)=\phi(x)_{ass}=C\,x_1x_2x_3\,\left(log \frac{q^2}{\Lambda^2}\right)^{-2/3\beta}
\end{equation}
Thus we find $C\simeq 0.13$ GeV$^4$. This value is consistent with the one of Eq.(\ref{cwave}), and shows that our simple approach to the proton form factor is quite sensible.

%------------------------------------------------------------------
%O paragrafo que foi colocado acima esta muito bom. A questao eh: isto eh uma copia grande ou pequena
%do que foi feito antes? Se a literatura anterior deixa isto muito evidente o paragrafo devera ser retirado
%e soh deverah ser informado o valor de C obtido desta forma.
%----------------------------------------------------------------
In conclusion, we introduced in the proton-proton elastic scattering calculation within the LN model the asymptotic 
behavior of the QCD proton form factor, supplied by a prescription about its infrared behavior in terms
of a dynamically generated gluon mass. The high energy behavior of the $pp$ diffractive cross section is determined as a function of the factor $C$, which is related to the qqq wave function at the origin, and turns out to be
obtained as a function of the gluon mass. We obtained $C= 0.15$ GeV$^4$ after fitting the elastic pp scattering data with $m_g= 460$ MeV for $\Lambda = 300$ MeV. Our result provide a desirable connection of the data with the
perturbative calculation of the proton form factor. Besides that we believe that the form factor expression
presented in Eq.(\ref{fullff}) can be further studied in different processes. Finally, the approximations that
we have performed are related to the poor knowledge of the gluon propagator and coupling constant
infrared behavior, and any improvement in the calculation could be only performed after an equivalent
improvement in the expression of the QCD infrared Green functions.

%%%%%%%%%%%%%%%%%%%%%%%%%%%%%%%%%%%%%%%%%%%%%%%%%%%%%%%%%%%%%%%%%%%%

\section*{Acknowledgments}

We are grateful to F.\ Navarra for useful discussions.
This research was supported in part by 
the Conselho Nacional de Desenvolvimento Cient\'{\i}fico e Tecnol\'ogico (CNPq)
(AAN, CZ),  Funda\c c\~ao de Amparo \`a Pesquisa do
Estado de S\~ao Paulo (FAPESP) (FC).

\end{document}